# Coherent Dual Comb Spectroscopy at High Signal to Noise

I. Coddington, W. C. Swann, N. R. Newbury

*National Institute of Standards and Technology, 325 Broadway, Boulder, Colorado 80305*

**Abstract**

Two coherent frequency combs are used to measure the full complex response of a sample in a configuration analogous to a dispersive Fourier transform spectrometer, infrared time domain spectrometer, or a multiheterodyne laser spectrometer. This dual comb spectrometer retains the frequency accuracy and resolution of the reference underlying the stabilized combs. We discuss the specific design of our coherent dual-comb spectrometer and demonstrate the potential of this technique by measuring the overtone vibration of hydrogen cyanide, centered at 194 THz (1545 nm). We measure the fully normalized, complex response of the gas over a 9 THz bandwidth at 220 MHz frequency resolution yielding 41,000 resolution elements. The average spectral signal-to-noise ratio (SNR) over the 9 THz bandwidth is 2,500 for both the magnitude and phase of the measured spectral response and the peak SNR is 4,000. This peak SNR corresponds to a fractional absorption sensitivity of 0.05% and a phase sensitivity of 250 microradians. As the spectral coverage of combs expands, coherent dual-comb spectroscopy could provide high frequency accuracy and resolution measurements of a complex sample response across a range of spectral regions.

**I. Introduction**

Fourier transform spectroscopy (FTS or FTIR) has long been a workhorse system for research and industry. Over 50 years since its inception, FTS remains relevant and continues to find new applications, in part, because of its continual refinement. The recent advances in spectroscopy using dual frequency combs offers an interesting new approach to FTS[1-9]. This dual-comb spectroscopy approach can also be viewed as a form of infrared time-domain spectroscopy (TDS) analogous to THz TDS [10-12], or as a massively parallel multi-heterodyne laser spectrometer [13].

Dual-comb spectroscopy has a number of advantages compared with conventional FTS. First and foremost is frequency accuracy. With the frequency combs stabilized to a known frequency reference, one can easily know the frequency of each tooth to an accuracy of one part in $10^{10}$-$10^{12}$ (and higher than a part in $10^{17}$ has been demonstrated [14]). Assuming the comb teeth can be isolated with high fidelity one achieves a 2-8 order of magnitude improvement in accuracy over a state-of-the-art FTS [15]. Second, since the frequency resolution is set by the comb repetition rates, rather than a scan distance as in FTS, high frequency resolution is achieved in a relatively compact, easily aligned system with no moving parts and without significant loss in signal-to-noise ratio (SNR) . At typical comb repetition rates, a dual comb spectrometer can easily reach 50-100 MHz resolution which is a significant engineering feat in traditional FTS. Moreover, it can do so with a reasonably high data acquisition rate, limited only by the Nyquist sampling limits discussed later. Third, as with laser spectroscopy, the single spatial mode output of the sources permits long interaction lengths.

Dual comb spectrometers do suffer from some disadvantages. First, the realization of the high frequency accuracy and resolution requires stabilized, phase-coherent frequency combs. Second, as with traditional FTS, the SNR on a single scan will be limited by dynamic range constraints, and an appropriate design is needed to achieve high SNR [16]. Third, while the spectral coverage of comb-based systems has been demonstrated over spectral spans greater than 14 THz [6], and continues to increase, it is and will remain limited compared to conventional FTS which can, in various configurations, cover the THz to the UV. Therefore, while dual comb-based spectrometers will not supplant conventional FTS, they provide an intriguing new solution for applications requiring high resolution, highly accurate measurements of a sample's complex response over spectral regions accessible to combs.

In earlier work, we demonstrated a coherent dual-comb spectrometer [6] and, more recently, have significantly improved its performance [9]. In this paper, we provide a detailed description of the setup and much more data on the system performance. This work describes the highest number of resolved elements and SNR demonstrated for a dual-comb spectrometer. A comparison of this system with conventional high-resolution FTS devices is difficult since it the two systems scale differently with sample length, resolution, spectral band etc. However, the quality factor (defined as the product of resolution elements and SNR [17]) achieved at 1550 nm and 200 MHz resolution is comparable (within 50%) to high-resolution FTS data in the same spectral region [18, 19] (scaled to the same resolution assuming an SNR proportional to the square of the spectral resolution [20]). While the demonstration here is still limited to a relatively narrow, 9 THz, spectral region, the basic approach outlined here should work as well with much broader comb sources and, more importantly, with comb sources further out in the mid and long-wave infrared spectral regions, as they are developed [21, 22].

The organization of this paper is as follows. Section II gives a general overview of coherent dual-comb spectroscopy with a discussion of the important attributes of frequency accuracy, resolution, normalization and sensitivity. Section III gives a detailed description of our system including the phase-locking of the comb sources and coherent real-time averaging. Section IV presents results in the time, frequency and joint time-frequency domains, and discusses potential system improvements. Finally, Section V concludes.

## II. Overview of dual-comb spectrometry
### II.A Frequency combs and spectroscopy

An ideal modelocked laser emits a train of pulses in the time domain at a constant repetition rate and with a constant coherent carrier extending across the pulses [23, 24]. In the frequency domain, this output corresponds to a frequency comb of evenly spaced "teeth" with the frequency of the $n^{th}$ tooth given by $\nu_n = f_0 + nf_r$ where $f_r$ is the pulse repetition rate and $f_0$ is an offset frequency common to all comb teeth. Since these sources combine wide spectral coverage with well defined discrete frequencies, they can provide a highly calibrated source for spectroscopy. However, there are at least two basic problems. First, such an ideal modelocked laser does not exist, since any real laser will suffer from noise. Therefore the comb must be actively stabilized or monitored [7] to remove noise in both the carrier frequency and repetition rate. Fortunately, in the case of passively modelocked lasers this requires only stabilizing, or monitoring, two degrees of freedom of the comb; the frequencies of all the comb teeth are then known. There are many methods for stabilizing the frequency comb, and the methods appropriate for measuring optical clocks are not necessarily optimal for broadband spectroscopy as discussed in Section III.B. Second, to take full advantage of the frequency accuracy and resolution of this comb, one must correctly and individually resolve each comb tooth in the final spectrum. This can be done with a high resolution spectral disperser [25, 26]. Combs have also been combined with FTS instruments with GHz resolutions [27-30] and a recent results at higher resolution [31] come very close to allowing one to resolve individual comb teeth. Alternatively, one can use the dual comb system provided sufficient coherence is maintained between the two combs, as is discussed here.

If lower resolution and accuracy is sufficient so that one does not need to resolve individual comb lines, one can still exploit the coherence of a comb by coupling it with a resonant cavity for increased sensitivity [8, 32-34].

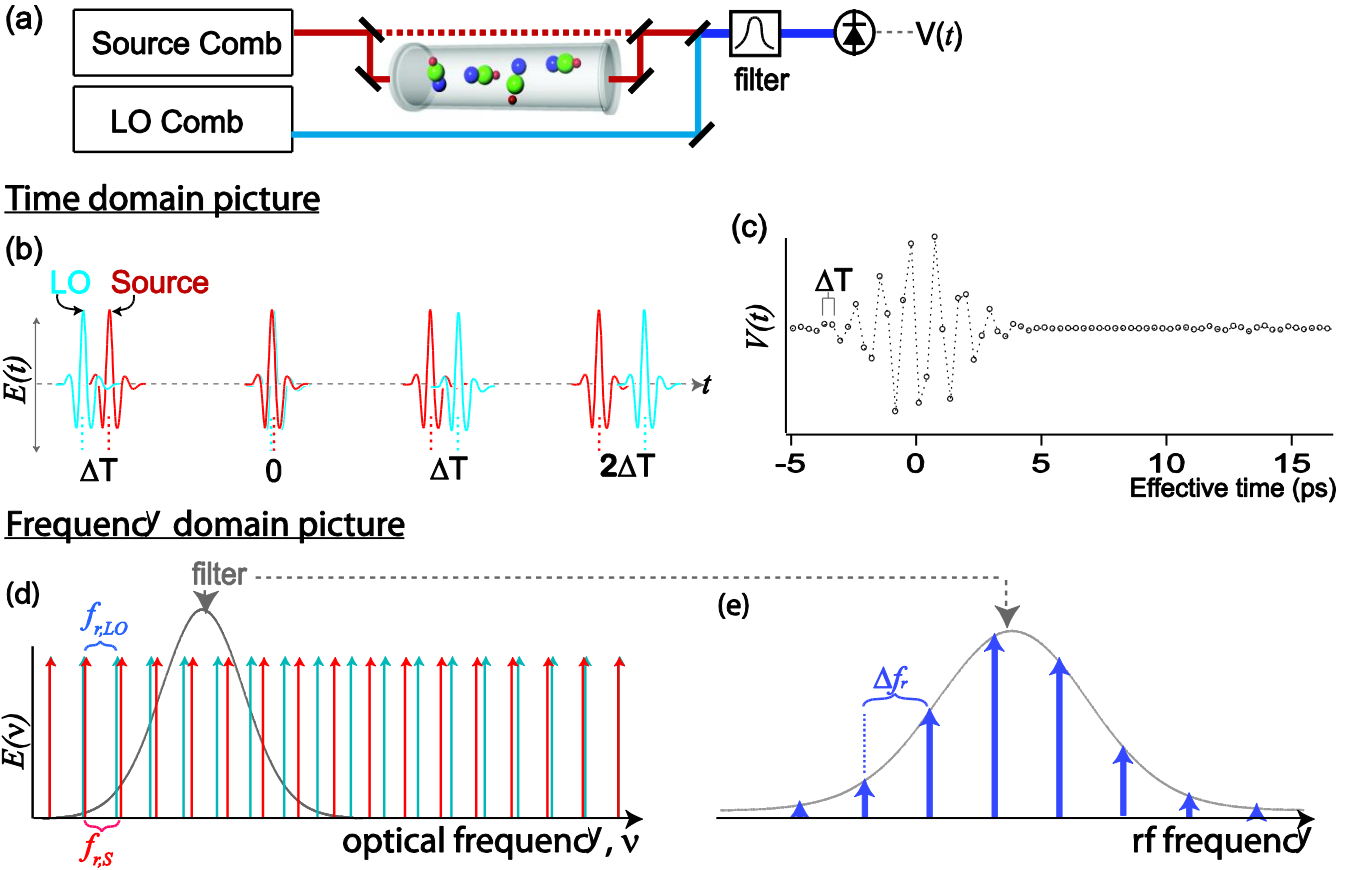

Fig. 1. (Color online) Coherent dual-comb spectroscopy relies on two combs with repetition rates $f_{r,S}$ and $f_{r,LO}$ that differ by $\Delta f_r$. (a) Simplified picture of a dual-comb spectrometer. The pulse train from a source comb passes through the sample (and is possibly split as well to pass around the sample to yield a time-multiplexed normalization signal). It is then combined with a Local Oscillator (LO) pulse train. Depending on the configuration, a tunable bandpass filter is employed to satisfy Nyquist conditions (see text). The heterodyne signal between the source and LO is detected and digitized to yield the complex gas sample response. In this configuration where the LO bypasses the sample, both the phase and magnitude of the sample response are measured. (b) Time-domain description of the coherent dual-comb spectrometer. The source and LO emit pulse trains in time. Since they have different repetition rates, the relative source and LO pulse timing increments by $\Delta T = \Delta f_r / (f_{r,S} f_{r,LO})$ with each sequential pulse. (c) Actual digitized data of the photodetector voltage corresponding to the overlap of the source and LO pulses. Each data point corresponds to the product of the source and LO pulses, integrated over the pulse duration. The detected signal is plotted verses effective time by assigning the data point spacing to be exactly $\Delta T$. Note that the data is equivalent to an interferogram, or cross-correlation, formed as pulses from the source and LO pass through each other. The large "centerburst" corresponds to the simultaneous arrival of the LO and source pulses. Also visible at 14 ps is a weak ringing containing absorption information from an HCN gas sample. (d) The equivalent frequency domain description of the coherent dual-comb spectrometer [1, 2]. In the frequency domain each comb creates an array of discrete teeth. The two combs each having slightly different repetition rates (tooth spacing) beat together to yield a third comb (e) in the rf with spacing $\Delta f_r = f_{r,S} - f_{r,LO}$. This rf comb is related to (c) simply by a Fourier transform and rescaling of the frequency axis. Note that for our experimental conditions the rf comb in (e) would contain ~4000 teeth.

## II.B Coherent Dual-Comb Spectroscopy

As shown in Fig. 1, the basic approach of dual comb spectroscopy is to interfere two combs with repetition periods differing slightly by $\Delta f_r$. If the two combs are combined before the sample, the system is analogous to traditional FTS [1, 2, 7] and yields the intensity absorption spectrum only. Alternatively, as is done here and in [3, 6, 8], the source comb passes through the sample and afterwards be combined with a local oscillator (LO) comb. This approach is analogous to asymmetric or dispersive Fourier transform spectroscopy (DFTS) [35, 17, 36, 37] and yields both the magnitude and phase spectrum of the sample. Note that for the single spatial mode output of combs, alignment issues are potentially less troublesome than in DFTS.

The operation of the dual comb spectrometer can be viewed in either the time domain or frequency domain (see Fig. 1d and 1e). In the time domain, the pulses from each comb source overlap on the photo detector at varying time delays, essentially creating a virtual scanning interferometer as in asynchronous optical sampling [10, 38]. The resulting cross-correlation of the source and LO pulses is analogous to a conventional interferogram in DFTS. An example interferogram is shown in Fig. 2. It has a centerburst followed by a trailing electric field, which, in the case of a molecular gas sample considered here, is just the free-induction decay (FID) of the molecules. A Fourier transform of this interferogram yields the spectral response of the gas and, since the interferogram is single-sided, one obtains both magnitude and phase. Because the pulse trains are periodic and repeatedly move through each other, there is no delay between interferograms. Rather a new interferogram begins to form as soon as the previous is finished. It is worth clarifying that the down sampling nature of this experiment introduces two time scales, which we will refer to as lab time and effective time. Lab time is the time taken to construct an interferogram and is characterized by a point spacing equal to the comb repetition period. Effective time corresponds to the time scale of a single reconstructed pulse and is characterized by the much smaller time step equal to the difference in comb repetition periods, given by $\Delta T = \Delta f_r / f_r^2$, where $f_r$ is the comb repetition rate (see Fig. 1b, 1c). Nyquist sampling limits require that the instantaneous optical bandwidth of the combs satisfy the relationship $\Delta \nu_{comb} < (2\Delta T)^{-1}$. Larger optical bandwidths can be measured by sequential

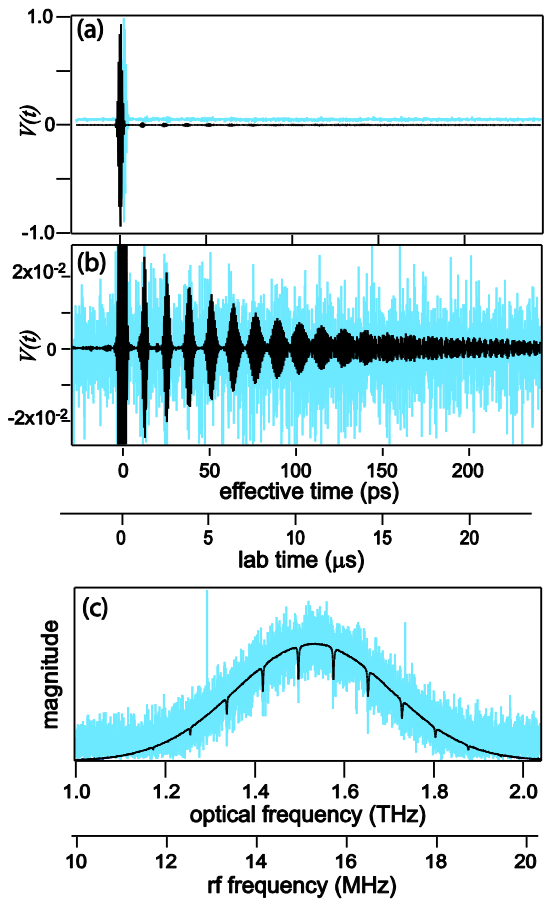

Fig. 2. (Color online) Example data traces illustrating the basic time domain signal and the ability of coherent signal averaging to improve signal-to-noise ratio (SNR). (a) The measured time-domain data (interferograms) for a sample of 25 Torr of hydrogen cyanide (HCN) gas around 1539 nm with a spectral bandwidth set by the tunable filter of Fig. 1a (filter used here is 350 GHz). For these data $\Delta f_r$=1 kHz, thus a single interferogram (blue trace) is acquired in a laboratory time of 1 ms (only the central ~ 8 μs is shown here). The time axis is also given in effective time, defined by the pulse-to-pulse time offset, $\Delta T$, between the signal and LO pulses. The strong signal at zero time, which has been normalized to unity, corresponds to the overlap of the LO and unperturbed source pulse. The effect of the gas is to generate a free-induction decay signal that extends to longer times. For the single trace (blue), this FID tail is barely above the noise level. However, the coherently averaged interferogram (black, offset for clarity-1000 averages shown) has greatly improved SNR. (b) Same data with the vertical axis expanded 25 times. The averaged data clearly resolves the free induction decay (FID) signal. This FID of the molecules appears as a pulses due to the "rotational recurrences" [57, 58], "rotational revivals", or "commensurate echoes" [11] as the rotating molecules rephase [9]. (c) The complex frequency-domain response is generated by a Fourier transform of the averaged trace. Spectral response (magnitude only) is plotted here for a single trace (blue) and 1000 averages (black). The rf frequency axis corresponds to the lab time axis and represents the actual frequency of the digitized signal. The optical frequency axis corresponds to the effective time axis and shows the actual optical frequency relative to the 1560 nm cw reference laser as discussed in Section II.E and III.D

measurements at different center frequencies, as done here.

In the frequency domain (Fig. 1d), one can view the system as a massively parallel laser heterodyne spectrometer. The phase and magnitude of each tooth of the source comb is measured through its heterodyne beat against one tooth of the LO comb. In order to insure a one-to-one mapping of the rf comb to the optical teeth, we require $\Delta \nu_{comb}/f_r < f_r/(2\Delta f_r)$, which is equivalent to the Nyquist sampling condition, $\Delta \nu_{comb} < (2\Delta T)^{-1}$ arrived at in the time-domain picture. Based on this Nyquist sampling constraints, for a given comb repetition rate and optical bandwidth, the minimum required time to acquire a spectrum equals $\Delta f_r^{-1}$. More rapid acquisition times, quoted for example in Ref. [8], refer to the apodization window applied to the interferogram, which nevertheless takes an acquisition time $\Delta f_r^{-1}$.

With this simple picture, several important questions are immediately apparent: how to normalize the signal to remove the distortions from the LO and source spectra, how to achieve high signal-to-noise ratio (SNR) at data rates of $f_r$~100 MHz or more (i.e. much higher than conventional FTS), and what are the critical requirements on the frequency combs.

**II.C Normalization**

The source light that passes through the sample is only measured with respect to the LO comb light and both can contain significant spectral structure in addition to etalon effects occurring in the beam path. Moreover this structure can vary with time. Without a rapid normalization scheme to remove these distortions, the effective SNR of the trace will be completely limited by background structure. In Ref. [6], we accomplished this with a second interferometer. Here and in Ref. [9], we have chosen instead to use a portion of the available interferogram time window by time-multiplexing a signal and reference pulse [39] (see Fig. 3) allowing us to update the reference every 300-1000 μs ($\Delta f_r^{-1}$). This approach sacrifices the potential 100-MHz resolution set by the comb repetition rate for the ability to measure a fully normalized spectrum in a compact setup with a single detection channel. It is particularly well suited to the single-sided interferogram because of the larger available time window. Because the reference and signal pulses travel through nearly identical optics etaloning effects are also strongly suppressed, allowing for an extremely flat baseline. An additional advantage of normalizing the data on a sub-millisecond time scale is the suppression of $1/f$ noise on the laser sources.

**II.D Signal to Noise Ratio**

The single point detector shown in Fig. 1 has the benefit of simplicity, high frequency accuracy, resolution, and adaptability to the mid IR and far IR. However it suffers an intrinsic SNR limitation compared to other comb spectroscopy that uses array detection [32, 33, 25, 26] because of detector dynamic range, and broadband laser technical noise [16]. The high frequency resolution and accuracy of a dual comb spectrometer are not of particular use unless the spectrum is measured with sufficient SNR. One route to higher absolute signal is a longer interaction length using a cavity or multi-pass cell. Cavity-enhanced single comb spectroscopy has been well demonstrated [32, 33] and cavity-enhanced dual comb spectroscopy has been recently demonstrated at 1 micron [8].

Regardless of the signal strength, one would like as high a SNR as possible. The SNR can be improved by averaging over many interferograms [3] in analogy with co-adding of spectra from rapid-scanning FTS systems [15, 35, 17, 20]. However, given the high digitizer rates (set by the repetition rates of ~100 MHz), the acquisition of multiple interferograms will lead to very cumbersome data file sizes, and furthermore individual phase-correction of these interferograms would require unacceptably long processing times. A solution for both these problems is to implement real-time coherent averaging by forcing the interferograms to have identical phase and summing them in real time. This coherent averaging can easily reduce file sizes by a factor of $10^4$ or more assuming only the coherently summed interferograms are saved. Coherent averaging is possible only while the two combs remain phase coherent; for longer averaging times, the standard phase correction techniques still must be used. Roughly speaking, coherent averaging requires a relative linewidth between the combs equal to the inverse of the averaging time, *i.e.* a relative linewidth much narrower than $\Delta f_r$. More details on our implementation of coherent averaging are given in Section III.C and the resulting improvement in SNR is shown in Fig. 2. As a final note, the ultimate solution to achieving high sensitivity will require the use of multi-pass cells or cavity enhancement, coherent signal averaging, and possibly multiple detectors.

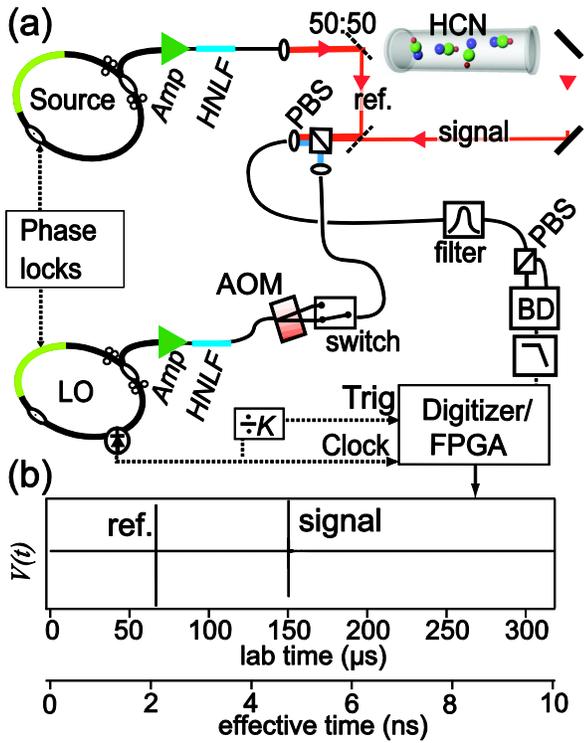

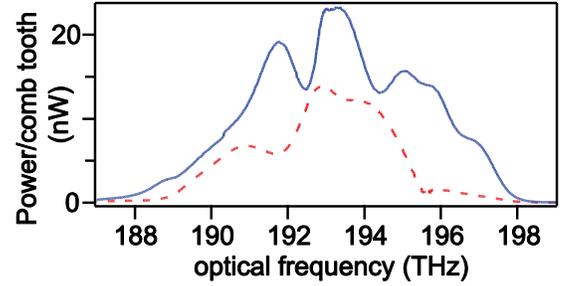

Fig. 3. (Color online) (a) Experimental setup. Two femtosecond erbium-doped fiber lasers are phase-locked together with a well-defined difference in repetition rates and phase-coherent optical carriers. Pulses from the source laser (or comb) are coupled into a free space section where a beamsplitter divides the source pulse into a signal pulse that passes through a cell containing $H^{13}C^{14}N$ gas and a reference pulse that circumvents it. Because of the different path length travelled by the signal and reference pulses, there is a 2.6 ns time separation between them. The signal and reference pulses are then recombined with each other, and then with the LO pulse train on a polarizing beam splitter (PBS). The combined pulses are coupled back into a single-mode fiber optic, filtered by a scanning bandpass optical filter (2 nm FWHM for data shown in Figs. 6-8) and detected on a balanced detector (BD). The additional AOM and switch in the LO path are used to select the offset frequency of the LO such that the detected rf multi-heterodyne signal between the LO and source combs does not fall at baseband or Nyquist (see Section III.A and Fig. 1b). (b) Detected voltage signal, with the peak normalized to one, corresponding to the overlap of the LO pulse with either the reference or signal pulse. Voltages are digitized synchronously with the LO pulses. The reference pulse arrives 2.6 ns sooner than the signal pulse and the separate arrival of the two is seen in the two separate peaks in the detected voltage. In the terminology of conventional FTS, these two peaks correspond to two separate centerbursts in the interferogram. The time axis is given in both laboratory time and effective time for a $\Delta f_r = 3.14\,\text{kHz}$. The ~320 microseconds duration in laboratory time corresponds to the time to acquire the $K$~32,000 LO pulses, spaced by ~ 10 nsec. The ~ 10 nsec duration in effective time corresponds to the same $K$~32,000 LO pulses with a point spacing set to equal the $\Delta T$~ 300 fsec time offset between the LO and source pulses. The data trace shown corresponds to actual data, averaged over $10^4$ individual traces using coherent real-time averaging, as discussed in the text.

Fig. 4. (Color online) Spectra of the two frequency combs measured just prior to the optical band pass filter. The LO comb is shown in blue (solid) and source comb in red (dashed).

## II.E Frequency comb stabilization

The basic picture of Fig. 1 assumes ideal source and LO frequency combs, however it is the real-world spectral and noise properties of the combs and their associated stabilization schemes that determine the frequency accuracy, resolution (and spectral coverage) of the spectrometer.

A frequency comb by itself has no inherent accuracy, as the free-running comb's two degrees of freedom can drift randomly; frequency accuracy is only obtained if the comb is either locked to or monitored against some pair of absolute references. For example, if one tooth of the comb is locked to an underlying cw reference laser, as is done here, the frequency of each tooth can then be inferred from this cw reference laser and the counted repetition rate. Without an underlying reference the accuracy of the must be derived from a known absorption line. The resulting accuracy is typically several orders of magnitude worse than is possible with stabilized combs.

The frequency resolution is in principle limited by the comb repetition rate (equivalent to the length of a mechanical arm as in FTS). However, a resolution equal to the repetition rate can only be achieved if there is no apodization and if the residual phase noise between the combs is low. As seen in Fig. 1(e), a resolution equal to the comb repetition rate requires a residual comb linewidth of $\Delta f_r$ [6]. (The residual linewidth is defined as the measured linewidth between teeth of the two combs, the absolute linewidths can be much larger). Otherwise, comb teeth in Fig. 1(e) will overlap ruining the one-to-one correspondence between the optical comb lines and rf comb lines, and limiting resolution to >$f_r$.

Residual linewidths below $\Delta f_r$ can enable coherent averaging, as described above, and can also allow a resolution below $f_r$. Strictly speaking, the system only samples at the discrete comb tooth frequencies, but one can in principle shift the combs and repeat the measurements to acquire data at interleaved comb frequencies. However, since spectral linewidths narrower than the repetition rate will have a time domain structure that extends across multiple interferograms, achieving resolution below $f_r$ requires continuous (unapodized) acquisition and processing of multiple contiguous interferograms. This, in turn, requires mutual coherence between the combs across the multiple interferograms. In other words, to have a resolution $n$ time better than $f_r$ one must have residual linewidths below $\Delta f_r/n$ (a fairly strong requirement).

Within the context of the above discussion, we briefly review the multiple approaches to stabilizing the frequency combs that have been used in dual comb spectroscopy. The use of free-running combs, i.e. femtosecond lasers, is certainly the simplest method and was used in a pioneering demonstration of dual comb spectroscopy [1]; however such an approach provides no absolute frequency accuracy, has limited frequency resolution and coherent averaging is challenging (although coherent averaging has been achieved in the far-IR where carrier phase is less of a problem [3]). In an intermediate approach, the carrier frequency of each comb is phase-locked to a common reference cw laser, and the pulse repetition rates are monitored [40, 41]. One can also monitor the carrier frequency of the free-running combs and correct the signal in post processing [7]. This approach removes the main source of drift – the carrier phase – and can provide high frequency accuracy if the reference laser frequency is known. To further enhance the coherence between the combs both combs can be fully phase-locked to a pair of calibrated cw reference lasers (the approach taken here). The accuracy of the combs is set by the cw reference laser frequency, which is measured with a self-referenced comb against a Hydrogen maser and is accurate to ~ 10 kHz (limited by drift in the cw reference between measurements). This accuracy generally far exceeds the statistical uncertainty in the line center of a Doppler- broadened line (even for the high SNR signals demonstrated here) and a looser cw reference with ~MHz accuracy locked to a simpler cavity, molecular line or even fiber loop [42] would suffice. With our approach, the phase locks are sufficiently tight such that the residual linewidths are ~0.3 Hz and allow for long coherent averaging times of 3 seconds and a corresponding improvement in SNR.

Finally if high phase stability was needed over a very broad bandwidth one could self-reference the comb using the standard carrier envelope stabilization techniques in combination with a lock to a single optical reference [43-46].

**III Experimental System**
**III.A. Optical layout**

Fig. 3 shows a more detailed picture of our setup. The two comb sources are erbium-doped femtosecond fiber lasers [47-50] with repetition rates $f_{r,LO} \approx f_{r,S} \approx 100\,\text{MHz}$ that differ by $\Delta f_r = 3.14\,\text{kHz}$ for most of the data shown here and are centered around 1560 nm. $\Delta f_r$ yields the effective time step $\Delta T = \Delta f_r / (f_{r,S} f_{r,LO})$ between the source and LO pulses (Fig. 1). The comb outputs are amplified to 20 mW in a short erbium-doped fiber amplifier which also provides a small amount of spectral broadening to ~ 9 THz as shown in Fig. 4. In Ref. [6], a nonlinear fiber was used to spectrally broaden the output but was not needed here. A splitter after the amplifier (not shown in Fig. 3) directs a portion of the light to the phase-locking setup described in the next section. The remainder of the source pulse train is directed through the interferometer signal and reference paths and polarization multiplexed with the LO pulse train. The combined reference and signal then pass through a tunable grating monochromator which filters the light prior to detection and digitization. For data shown in Fig. 6-8 the filter has 2 nm (200 GHz) full-width half-maximum (FWHM) Gaussian profile. Data taken with different filter center wavelengths are separately processed, as described later, and then coherently stitched together to generate the total spectrum.

The use of an optical filter is not absolutely necessary [1, 2, 7, 8] and might at first seem unwanted. On one hand the filter reduces the power on the detector and requires multiple measurements to reconstruct the spectrum. However, in a shot noise limited system this additional measurement time is exactly offset by the lower noise of a reduced spectrum. Moreover, in the dynamic-range limit (set by detectors or digitizers), filtering the spectrum can lead to an overall increase in SNR [16]. In our case, the system is dynamic range limited near the center of the spectrum and the filter improves the SNR. A more complete discussion of the SNR limits is given in reference [16].

More significantly, to spectrally filter is to simultaneously meet Nyquist sampling conditions and suppress $1/f$ noise. Specifically, the filtered comb bandwidth ($\Delta\nu_{comb}$) must be limited to $\Delta\nu_{comb} \leq 1/(2\Delta T) = f_{r,S} f_{r,LO}/(2\Delta f_r)$, to avoid aliasing, as stated in Section II.B. Thus a broad spectral bandwidth requires a low value for the difference in comb repetition rates, $\Delta f_r$. However, the interferogram updates at $\Delta f_r$ and will be degraded by $1/f$ noise in the comb spectra at low values of $\Delta f_r$. The use of the filter allows us to maintain the high scan speed ($\Delta f_r$) by limiting the instantaneous spectral bandwidth, and later recover the entire spectrum through coherent stitching.

After the filter, the orthogonally polarized signal and LO pulses are mixed with a fiber-coupled polarizing beam combiner oriented at $45^\circ$ with respect to the laser polarization states. Polarization paddles are used to fine tune the balancing of light levels. The two outputs of the beam combiner are incident on a 110 MHz commercial balanced detector. Balanced detection suppresses the strong homodyne signal from the individual pulses and allows maximization of the dynamic range available for the heterodyning between the pulses. The output of the detector is low pass filtered at 50 MHz to avoid aliasing of signals above the Nyquist frequency and to relax the timing between the digitizer clock and the pulse arrival on the photodiode. The detector signal is then digitized by a 100 MHz 12-bit digitizer. The detector and internal amplifiers have a 1 mV noise floor and begin to show saturation behavior at $\pm 500\,\text{mV}$ giving us our 500:1 dynamic range limited SNR. A better detector would improve the dynamic range to ~2000:1, the limit set by the 10.3 effective number of bits of our 100-MHz digitizer [15].

As shown in Fig. 3, the LO passes through a 120 MHz acousto-optic modulator (AOM) and fiber switch to select either the first (shifted) or zeroth (unshifted) order beam from the AOM. There are two benefits to the AOM. First the LO power can be adjusted via the AOM's rf power to avoid photodetector saturation as the optical filter is tuned from high to low power regions of the comb spectra shown in Fig. 4. Secondly, at certain optical filter positions the rf beat between the combs falls near the Nyquist frequency ($f_{r,LO}/2$) or near 0 Hz. At either of these frequencies, aliasing effects lead to spurious data. We avoid these regions by shifting the rf interferogram using the first-order deflected beam thus shifting the LO by $\sim f_{r,S} + f_{r,S}/4$ ($\sim 125\,\text{MHz}$). Maintaining carrier phase of the interferogram during coherent averaging (see Section III.C) does require the AOM be driven at an integer multiple of $\Delta f_r$, but this shifting is otherwise straightforward. One could suppress these reflections about Nyquist with IQ detection however it is difficult to realize more than 20 dB of suppression from an optical IQ detector and $1/f$ noise would remain near 0 Hz. One could also adjust lock points.

**III.B Phase Locking of the source and LO combs**

Fig. 5a shows the basic method used here to phase-lock the two Er fiber combs together. A pair of comb teeth from each comb are locked to two cw reference lasers at 1535 nm and 1560 nm respectively. With this setup, in the frequency domain the source and LO combs exactly overlap every $K$ and $K+1$ modes respectively. In the time domain, the LO and source pulses exactly overlap every $K$ pulses and $K+1$ pulses respectively. $K$ sets both the number of samples in a single interferogram and the scaling between lab time and effective time (*e.g.* Fig 3b axes). For most data shown here, $K = 31834$. The integer length of the interferogram greatly simplifies coherent signal averaging as discussed below.

Both cw reference lasers are stabilized to the same high-finesse cavity via a Pound Drever Hall lock (PDH) [51] as shown in Fig. 5d. The PDH lock is implemented using almost all fiber-optic components, for the first time to the authors' knowledge. The absolute frequency of the 1560 nm cavity-stabilized laser is measured using a separate, fully self-referenced frequency comb linked to a hydrogen maser [14, 45]; this absolute frequency was observed to drift by ~ 10 - 30 kHz over days. This level of absolute frequency accuracy is excessive for many spectroscopic applications and other less accurate stabilization methods for the cw reference lasers are certainly possible as note in Section II.E. In practice, we use the measured frequency of the 1560-nm laser and the measured repetition rates of the two combs, with respect to a Hydrogen maser, to calculate the optical frequency of the source and LO comb teeth. Our accuracy is linked directly back to the Hydrogen maser in this fashion. Note that in this case, the 1535 nm laser does not need to be cavity-stabilized as well but it is convenient.

The two combs are phase-locked to the cw lasers in an identical fashion. The comb light is combined with the two cavity-stabilized cw lasers, spectrally filtered to two channels at 1535 nm and 1560 nm with a 1 nm width, and detected on two 100 MHz detectors, as shown in Fig. 5e. Each detector detects a beat signal between the comb and one cw laser. The two beat signals are then used in two separate servos to tightly phase-lock the combs. Coarsely speaking we use the 1560 nm light to stabilize the comb carrier frequencies and the 1535 nm light to stabilize the comb repetition rates (in truth these parameters are coupled). Because the optical bandwidth is relatively small compared to the carrier frequency, the carrier stabilization lock is the more critical of the two locks. For this lock, we divide the rf beat signal between the comb and 1560-nm laser by eight to increase the capture range and lock the divided signal to a 10 MHz reference by feeding back to an AOM with ~100 kHz bandwidth and a much slower, but higher dynamic range, intracavity PZT. The servo to the PZT will effectively remove much of the lower frequency noise on the repetition rate of the comb as well. However, to fully stabilize the repetition rate, we use the beat signal between the comb and the 1535 nm light, which is fed into a servo filter whose output is fed to the pump current for the laser cavity. By controlling the pump current, we control the repetition rate while roughly holding the carrier frequency fixed [52]. The feedback bandwidth is ~10 kHz limited by the laser dynamics combined with the erbium response [53]. The 10 kHz is sufficient, however, phase lead compensation could achieve higher bandwidths [44]. The resulting phase-locked rf signal and residual phase noise of the phase locks are given in Fig. 5b-c.

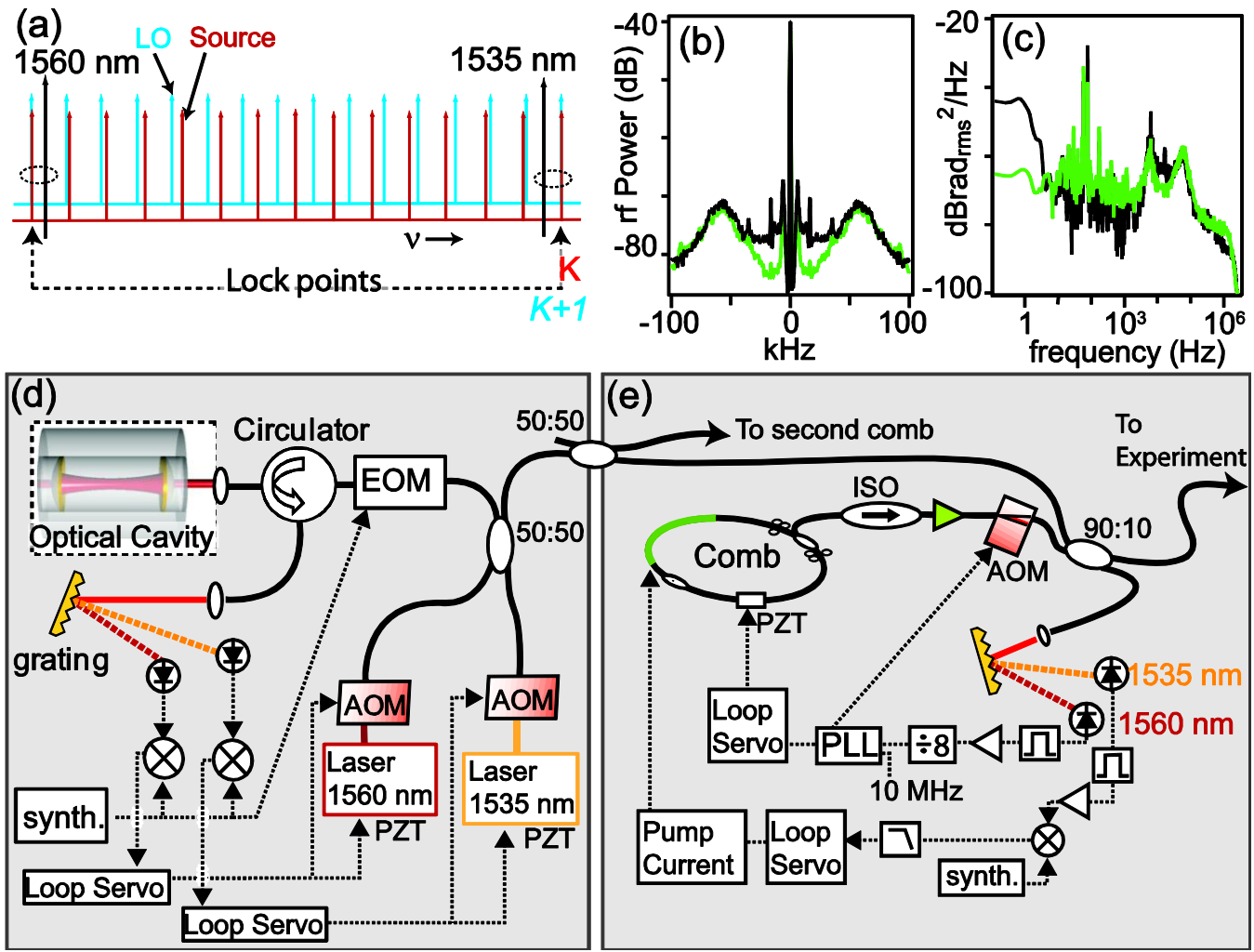

Fig. 5. (Color online) Setup used to stabilize the two frequency combs such that they are both phase-coherent with sub-Hertz residual linewidths and with a very well-defined difference in repetition rates, $\Delta f_r$. (a) The locking setup is most easily illustrated in the frequency domain picture. A pair of teeth from each comb is stabilized to a pair of cw reference lasers. The phase-locks use identical rf offsets for both combs. Repetition rates are chosen such that there are exactly $K+1$ LO comb teeth between the two points and $K$ for the source comb. As a consequence, in the time domain, the interferograms repeat precisely every $K$ LO pulses. (b) Measured RF power spectrum for the phase locked rf beat between the source comb tooth and the 1560 nm laser (black) and the 1535 nm laser (green) at a resolution bandwidth of 125 Hz. The rf spectra corresponding to the LO locks are similar. (c) Phase noise spectrum for the same phase locks as (b). The integrated residual phase jitter ranges from 0.27 to 0.47 rad (from 0.1 Hz to 3 MHz) for all four phase locks. (d) The setup used to stabilize the two cw reference lasers to a high-finesse optical cavity. A Pound-Drever-Hall cavity lock was used with a common fiber-optic path for both cw lasers, which were only separated to measure the final error signals. AOM- acousto-optic modulator, EOM- electro-optic modulator, synth.-tunable rf frequency synthesizer. (e) Setup used to phase-lock a single comb to the two cw reference lasers. Details are given in the text. PLL-Phase lock loop, ISO- optical isolator, PZT-piezoelectric transducer.

**III.C Data Acquisition**

The signal is digitized at the LO repetition rate to match the effective optical sampling of the LO pulses, which translates to a high sample rate. As discussed in Section II.D, to reach a high SNR requires signal-averaging of multiple interferograms just as in conventional FTS. However, the brute force approach of directly digitizing the signal for as long as possible [6] leads to massive data sets and makes deep averaging impractical. Therefore, instead we implement coherent averaging by phase-locking the two combs as described above such that each interferogram is exactly an integer number ($K$) of points with exactly the same phase [9]. In that case, we can simply sum successive interferograms in real time either onboard an FPGA or in software for a time approximately equal to the inverse of the relative linewidth between the combs. The requirement on the relative comb linewidth for coherent averaging is very demanding; we require linewidths $< \Delta f_r/N_A$, where $N_A$ is the number of coherently averaged interferograms.

For our system, we are able to coherently average for ~3 seconds corresponding to $N_A = 10,000$ interferograms and a ~0.3 Hz relative comb linewidth. The commercial digitizer must be triggered for each successive interferogram. We generated a trigger, accurate to below the 10ns clock period, by using a digital pulse counter to divide the LO clock signal by $K$ (Fig. 3). Our digitizer does require a small (~20 μs) dead time between triggers which leads to a slight loss in frequency resolution (~7%). At our 200 Mbyte per second sampling rate, a 600 MB data stream is reduced to 60 kB (or by $10^4$) with a corresponding decrease in processing time because the FFT is now performed on the shorter data set. For acquisition periods greater than 3 seconds, drifts in optical path lengths and noise in the phase locks leads to a relative carrier phase drift between combs. Therefore, we must first phase-correct the sequential 3-second coherently summed interferograms, based on the phase of the centerburst corresponding to the reference peak [39], and then sum them to generate an interferogram at longer averaging times. One can also phase correct more frequently to relax the requirements on the mutual coherence (residual linewidth) between combs but the processing load and data transfer from the digitizer will add additional experimental dead time. However, more sophisticated FPGA's could make this realtime correction practical.

**Section III.D Fourier Processing**

The first step of the data processing is to change the scaling of the data from laboratory time to effective time. Given the locking scheme used here, this scaling is straightforward. We simply set the sample point spacing to equal the effective time step $\Delta T = \Delta f_r / (f_{r,LO} f_{r,S}) = 1/(K f_{r,S})$. The three quantities $\Delta f_r$, $f_{r,LO}$, $f_{r,S}$, are measured with frequency counters.

The time multiplexing of reference and signal pulses discussed earlier yields the optical interferogram, $V(t_i)$ seen in Fig. 6(a), where $t_i$ is in effective time, with the two pulses separated by 2.6 ns and the signal peak is defined to be zero time. A "reference-only" time trace is generated by gating out all but the reference pulse, to form $V_R(t_i) = \langle V(-4.7 \text{ ns} < t_i < -1.3 \text{ nsec}) \rangle$ and zero-padded to a length $K$. Similarly, a "signal-only" time trace is generated by gating out all but the signal pulse to form $V_S(t_i) = \langle V(-1.3 \text{ nsec} < t_i < 4.6 \text{ nsec}) \rangle$, which is also zero-padded to $K$ points. To deconvolve the reference from the signal, the two signals are first Fourier transformed to yield their respective complex spectra (Fig. 6(b)). The ratio yields the complex molecular response, shown in Fig. 6(c), for the filter centered at $\lambda$ as $\tilde{H}_\lambda(\nu_k) = \tilde{V}_S(\nu_k)/\tilde{V}_R(\nu_k)$, where the tilde represent a Fourier transform evaluated on the optical frequency grid $\nu_k$ with a known optical offset $\nu_0$ discussed below. (The response is related to the linear susceptibility as discussed in Section IV.) Because the time trace is padded to $K$ points, the frequencies $\nu_k$ correspond precisely to comb teeth of the signal comb. The resolution of the experiment is set by the temporal length (*i.e.* apodization) of $V_S(t)$ giving a resolution of $1/4.6\,\text{ns} \approx 220\,\text{MHz}$. The reference pulse is more strongly apodized with less spectral resolution to limit the contribution of noise from the reference to the final deconvoluted signal. Generally the reference pulse window must only be broad enough to contain any etaloning features ($<1\,\text{ns}$).

The optical frequency offset, $\nu_0$, is calculated from the measured frequency of the 1560 nm cw laser, the known rf offsets in the phaselocking, the known AOM frequency shift if present, and the measured repetition rate. For example, for the unshifted LO comb, $\nu_0 = \nu_{1560} + f_{lock} + pKf_{r,S}$ where $\nu_{1560}$ is the measured frequency of the 1560-nm cw reference laser, $f_{lock}$ is the rf offset of the phase lock to the 1560 nm laser, and $p$ is an integer chosen such that $\nu_0$ is close to the optical bandpass filter center. The term $pKf_{r,S}$ reflects the ~1.5 THz Nyquist ambiguity resulting from the $\Delta T$ time step. In addition the sign of the frequency axis of spectra in Fig. 6(b) is changed from positive to negative as necessary. Assigning the negative sign and values for $p$ can be done with ~1 nm level knowledge of the filter position. Thus applying these shifts correctly requires only a coarsely calibrated optical filter.

Despite the normalization, some slow background wander will remain on $\tilde{H}_\lambda(\nu_k)$ due to multiplicative phase noise [16]. While small (~ 0.1% in magnitude or 1 mrad in phase), this wander can lead to ripple on the concatenated data. Therefore, we first remove it by separately fitting the magnitude and phase profiles of $\tilde{H}_\lambda(\nu_k)$ to a third-order polynomial over a range covering the FWHM of the optical filter. The magnitude is then normalized to unity by dividing by the magnitude fit, and the average phase is set to zero by subtracting the phase fit. Strong absorption features can throw off the fits, so we mask out features that deviate from the baseline by more than 2 %. For the phase profile it is sufficient to apply a 7 GHz box-smooth prior to fitting. Note that this filtering technique would have to be reconsidered if one were looking at very broadband spectral features.

The individual $\tilde{H}_\lambda(\nu_k)$ are then coherently stitched together, i.e., concatenated, to generate an overall response function covering the full frequency span and at a spacing $f_{r,S}$. Because the combs are phase-locked to stabilized cw lasers, the optical frequencies are stable over the measurement period and there are no issues concatenation. This processing involves only fast Fourier transforms, which are indeed fast if $K$ is factorable into small prime numbers.

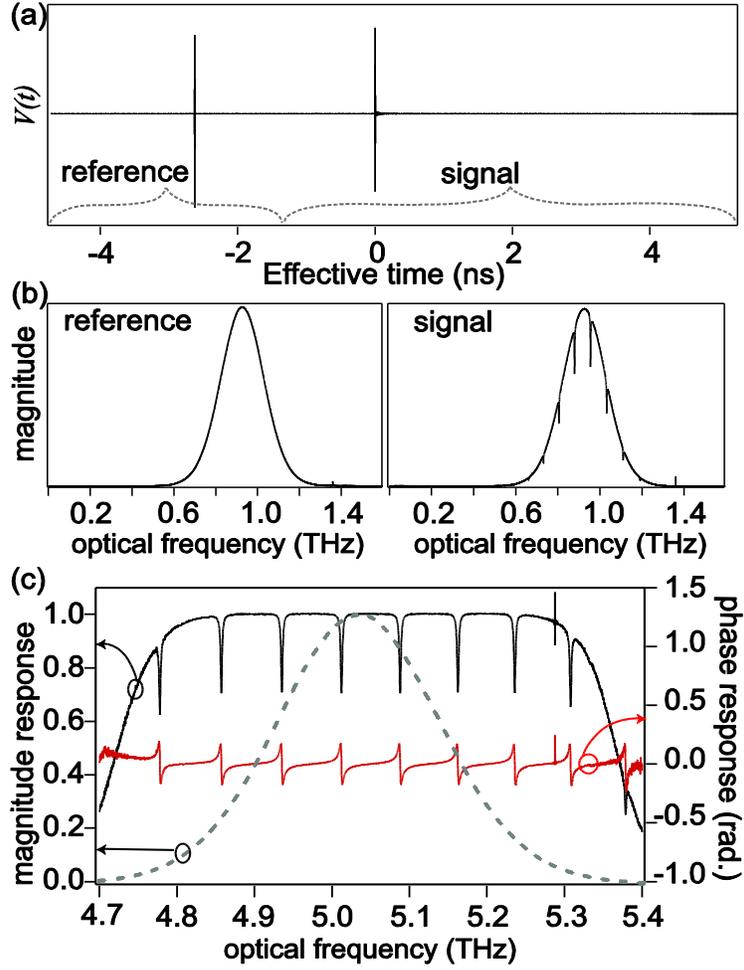

Fig. 6. (Color online) Data processing for the signal from a 15 cm long cell filled with 25 Torr of $H^{13}C^{14}N$ gas. (a) Time domain interferogram signal after 10,000 averages. The trace is divided into a reference and signal portion as shown. The reference corresponds to the cross-correlation of the LO pulses with the source pulse that circumvented the sample while the signal corresponds to the source pulses that passed through the sample. (See Fig. 3a). (b) The Fourier transform of reference, $\tilde{V}_R(v_k)$, and signal, $\tilde{V}_S(v_k)$, sub-traces (magnitude only, phase not shown). The signal shows the absorption of about 6 ro-vibratonal HCN lines. (c) Normalized response, $\tilde{H}_\lambda(v_k)$, in magnitude (black) and phase (red) calculated from the ratio of the complex, frequency-domain signal and reference of part (b), after shifting the frequency axis by the offset frequency, $v_0$. Dotted line shows the reference trace with the same offset. A spurious peak at 5.29 THz due to rf pick up is visible here but removed from the final data. When response traces are concatenated to cover the full comb spectral width, only the data within the filter FWHM are used.

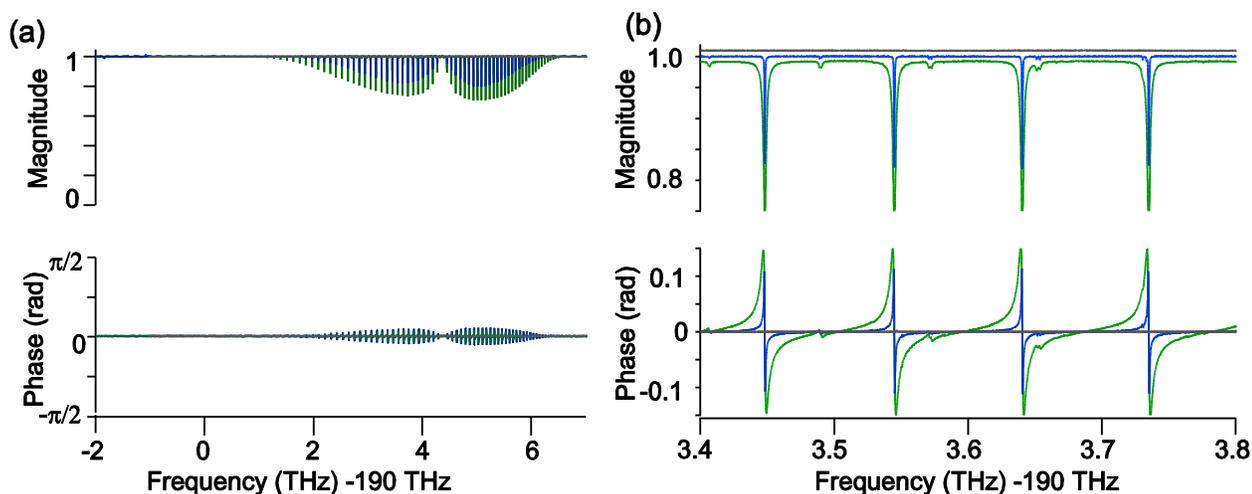

Fig. 7. (Color online) (a) The measured complex spectral response, $\tilde{H}(\nu)$, of the ro-vibrational lines corresponding to the first overtone H-C stretch vibration in HCN over a 9 THz bandwidth at 220 MHz resolution. The "p-branch" and "r-branch" are clearly visible. For each isolated ro-vibrational line, the magnitude corresponds to the standard gas absorption profile (divided by two since absorption is typically given for intensity) and the phase corresponds to the delay of the light due to the change in the index of refraction. Three different HCN gas pressures are shown: 25-Torr in a 15-cm long cell (green), 2.7-Torr in a 20-cm long cell (blue), and an empty cell (grey). Each trace is generated by concatenating the complex frequency response (*e.g.* Fig. 6c) measured at forty five different settings of the tunable 2 nm-wide optical bandpass filter. (b) Expanded view of the complex HCN spectrum. For the magnitude data, the three traces have been offset for clarity. For the higher-pressure cell (25 Torr), the magnitude has the Lorentzian profile expected from collisional line broadening with a corresponding "derivative"-like shape for the phase profile. For the lower-pressure cell (2.7 Torr), the magnitude has a gaussian profile expected from Doppler line broadening, again with corresponding "derivative"-like shape for the phase profile. The molecular hot bands can clearly be seen as much smaller magnitude and phase excursions. The combined benefits of extended averaging and normalization by the reference channel can be seen in the flatness of the baseline. The statistical noise averaged over the full spectrum is ~0.04% in magnitude and 400 μrad in phase. Over the central portion of the spectrum, it drops to ~0.025% in magnitude and 250 μrad in phase. These data were acquired with 60 seconds of averaging at each of the forty five positions of the tunable optical filter giving a total acquisition time of 2700 sec. Of course, the SNR scales with the square root of time so that a 270 sec total acquisition time reduces the SNR in magnitude and phase by a factor of only ~3.3.

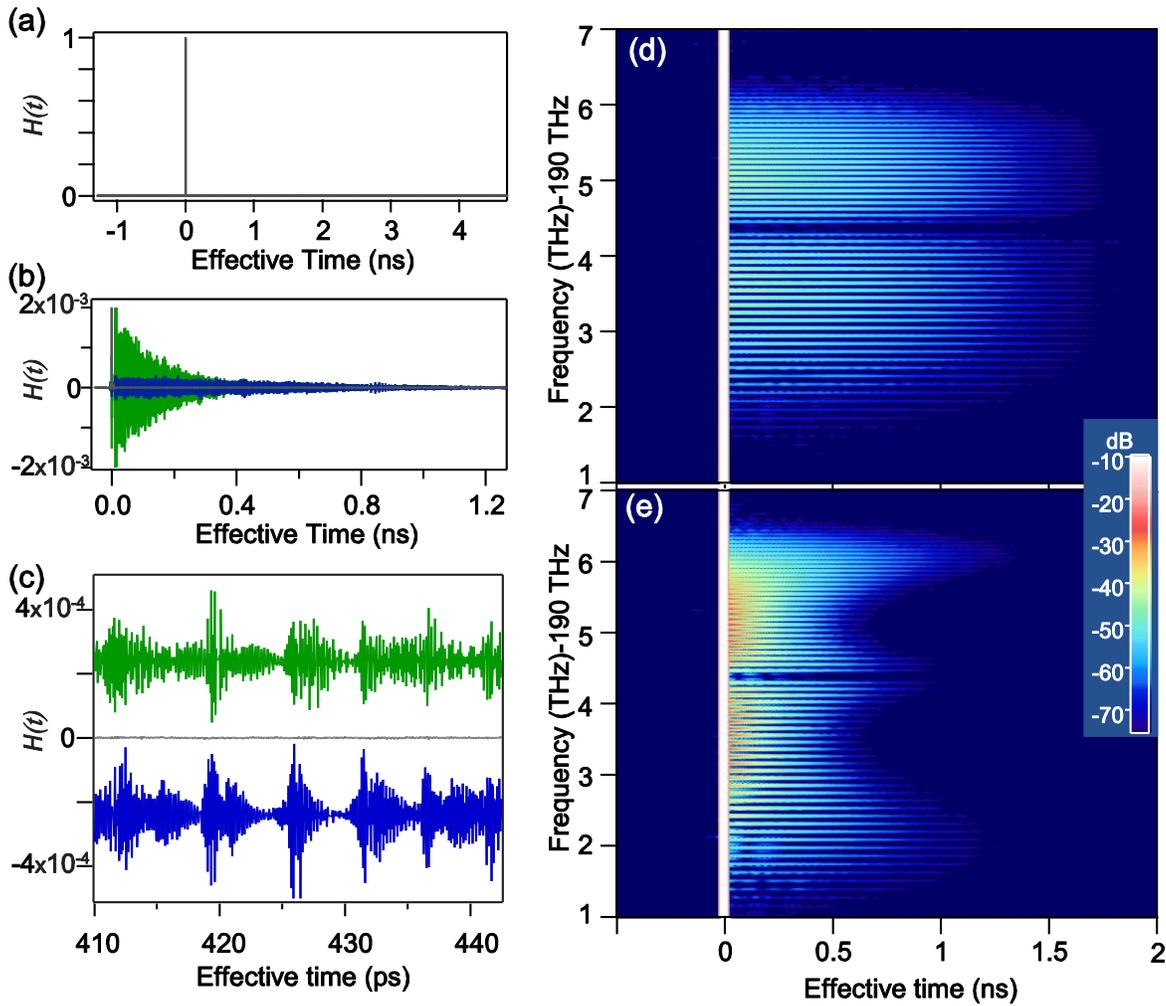

Fig. 8. (Color online) The reconstructed time-domain response and joint time-frequency domain response of HCN gas corresponding to the complex frequency response of Fig. 7. This different, but equivalent, view of the data allows for a different physical picture of the gas response as discussed in the text. (a) The total time-domain response over a 6 ns window at 55 fs point spacing ($9\,\text{THz}$ bandwidth) with the peak at $t=0$ normalized to unity. The time-domain SNR is $0.55\times10^6$. (b) Expanded view of the time-domain response for the 2.7-Torr cell (blue), 25-Torr cell (green) and empty cell (gray). While the empty cell trace is flat, the filled cells clearly show a tail from the FID of the excited HCN molecules. For the lower pressure, 2.7-Torr cell, 170 rotational recurrences are visible out beyond 1.8 ns. Because of causality, the FID appears only at positive times (where $t = 0$ corresponds to the arrival of the unperturbed excitation pulse). This single-sided interferogram is a direct consequence of the fact the LO pulse train passes around rather than through the HCN sample. The envelope of the overall FID signature decays more slowly for the lower pressure cell because of the much lower collisional decoherence rate. (c) Further expanded view where the 25-Torr and 2.7-Torr cell data are offset for clarity. The low noise level can be seen in the middle empty cell trace (gray). The complicated time-domain structure is a result of interferences between the different ro-vibrational levels. (d) Sonogram (short-time Fourier transform) of the 2.7 Torr data processed with 19 GHz frequency resolution and 52 ps time resolution. The bright vertical stripe at $t = 0$ corresponds to the arrival of the signal pulse and has been normalized to 0 dB. At this frequency resolution the trailing decay signal from each absorption line is clearly visible. At 2.7 Torr, decay is driven primarily by Doppler dephasing, allowing for relatively long decay times. (e) Sonogram under identical conditions as (d) but for the 25 Torr data. At this pressure collisions accelerate the decay time and collisional resonance [70] between like states is evident in the thermal distribution appearance to the decay times.

**Section IV: Results and Discussion**
**IV.A Sample response in the frequency, time, and joint time-frequency domains**

In Fig. 7, we present the measured frequency-domain response in terms of both transmission and phase shifts for the ro-vibrational band corresponding to the first overtone C-H stretch in hydrogen cyanide (HCN) for a 2.7-Torr cell and a 25-Torr cell, along with an empty reference cell. (Data for the 2.7 Torr cell of HCN were also given in Ref. [9].) The frequency domain response is given by $\tilde{H}(\nu) = 1 + 4\pi^2 ic^{-1}\nu\chi(\nu)L$, where $L$ is the length, $\chi$ the linear susceptibility of the sample, $c$ is the speed of light, and $\nu$ is the optical frequency. For weak gas absorption $\tilde{H}(\nu) \approx 1 - \alpha(\nu)L/2 + i\Delta k(\nu)L$, where $\alpha(\nu)$ is the usual absorption coefficient and $\Delta k(\nu)$ is the corresponding phase shift. For the case of simple, well separated absorption lines considered here, the absorption lines appear with the standard Voigt profile and the phase shift with the corresponding "derivative"-shape. (These are actually just the real and imaginary parts of the complex Voigt function or error function for complex arguments [54]). Note that this signal differs from the simple absorption spectrum, $\alpha(\nu)$, usually generated in grating spectrometers or typical double-sided FTS, because we retrieve both the magnitude, $\alpha(\nu)/2$, and the phase. Of course, using Kramers-Kronig relations one can in principle retrieve the phase from the magnitude spectrum; by measuring it directly we avoid any complications due to finite acquisition bandwidth.

In the time domain (see Fig. 8a-c), the signal is simply the Fourier transform of the above frequency-domain signal convolved with a sinc function (corresponding to the spectral bandwidth, $\Delta\nu$, of the normalized response) or
$$H(t) \approx e^{-i2\pi\nu_c t}\text{sinc}(\pi t \Delta\nu) \otimes \left[\delta(t) + 4\pi^2 ic^{-1}\nu_c \tilde{\chi}(t)L\right]$$
, where we approximate the Fourier transform of $\nu\chi(\nu)$ as $\nu_c\tilde{\chi}(t)$ where $\nu_c$ is the average carrier frequency. Physically, we are observing the forward scattered light or free induction decay of the excited molecules. The initial pulse excites a superposition of ro-vibrational states. These vibrating states act as dipole emitters. However, they quickly rotate out of phase with each other to give no coherent forward scattered signal. Because the rotation rates are quantized, they do rotate back in phase a short time later (set by the rotational constant), giving rise to another burst of coherent radiation in the forward direction, long after the excitation pulse has ended (see Fig. 2b or 8c). The mathematical description follows earlier work on two-level systems by Brewer and co-workers [55, 56], or the work on commensurate echoes in THz time-domain spectroscopy [11] or rotational recurrences in pump probe spectroscopy [57, 58].

Of course, there is no particular reason to view the signature in the purely time or frequency domains, and in Fig. 8d-e we show a joint time-frequency domain picture at two different gas pressures. In this picture, the band at zero time is the reconstructed initial pulse and the trailing signal at later times is the frequency-resolved free induction decay amplitude. Each HCN absorption line is (slightly counter-intuitively) evident as a trailing signal. The effect is due to the slowing of the light near the resonance which delays the transmitted light at that frequency. The overall signal decays as a result of collisions and Doppler effects. The collisions cause a phase-shift or de-excitation of the vibrating, rotating molecules so that they no longer add coherently. The Doppler shifts can be viewed as the molecules moving out of their original position so that their emitted dipole radiation no longer adds phase-coherently in the forward direction. Collisional effects dominate for the higher pressure cell and the effects of resonant collisions is evident in the faster relaxation of the more highly populated ro-vibrational states. Doppler relaxation dominates for the lower pressure cell.

The nominal signal-to-noise ratio will depend on whether it is quoted in the time or frequency (or joint time-frequency domains). The time-domain SNR is measured as the ratio of the peak of the interferogram to the standard deviation of the noise as measured from either the empty cell or, equivalently, before the centerburst. At a point spacing of 55 fs over the 6 ns time window, we obtain a time-domain SNR of

$$SNR_t = 1.1 \times 10^4 T^{1/2} \qquad (1)$$

as a function of total acquisition time, $T$. For the longest acquisition time of $T = 45$ minutes for the data of Fig. 8, we measure $SNR_t = 0.55 \times 10^6$. (Note that in calculating this SNR we use the noise near the centerburst, which is larger by $\sqrt{2}$ than the noise at longer time offsets due to the effects of deconvolution.) This noise carries over to the frequency domain, where we measure an average SNR across the 41,000 frequency elements that span the full 9 THz of

$$SNR_f = 50\,T^{1/2} \quad (2)$$

in radians or fractional magnitude change in the reconstructed spectrum. The peak SNR near the center of the spectrum is about twice as high or $100T^{1/2}$ over a 1 THz window.

If the time-domain noise is white, then the transformation from Eq. (1) to Eq. (2) is straightforward. Specifically, $SNR_f = SNR_t/\sqrt{M}$, where $M = 9\,\text{THz}/220\,\text{MHz} = 41,000$ is the number of resolved frequency points. Although the frequency noise is not completely white due to variations in the comb power across the spectrum, this relationship roughly holds because $SNR_t/\sqrt{M} = 54\,T^{-1/2}$ in agreement with Eq. (2). The SNR is primarily limited by the dynamic range of the system and agrees with the calculated SNR for our conditions and dynamic range using equations given in [16].

In terms of sensitivity for trace gas detection, the minimum detectable absorption $L\alpha_{\min}$ is often used as system metric. Eq. (2) gives the SNR for the magnitude, $L\alpha(\nu)/2$, so that one might consider twice that value as a measure of the sensitivity. However, this relation oversimplifies the situation for several reasons. First, the measurement is made at much higher resolution than a typical spectral linewidth; the signal across the entire line contributes to the overall sensitivity and there is a corresponding increase in sensitivity over Eq. (2) of a factor of $\sim \sqrt{\Delta\nu_L/\nu_{res}}$, where $\Delta\nu_L$ is the FWHM collisionally broadened linewidth and $\nu_{res}$ is the system resolution. (This is the SNR enhancement for smoothing the data to the resolution of the spectra of interest). Moreover, the measurement is made over multiple spectral lines and the signal across all the spectral lines contributes to the overall sensitivity. Therefore a matched filter yields a sensitivity that is actually improved by a factor of $\sqrt{(\pi/4)\sum_i \alpha_i^2 \Delta\nu_{Li}/(\alpha_0^2 \nu_{res})}$ where the $i^{\text{th}}$ line has peak absorption $\alpha_i$ and Lorentzian FWHM linewidth $\Delta\nu_{Li}$ and the sum is over all ro-vibrational lines [16]. With these extra factors, for HCN, the absorption sensitivity is improved over Eq. (2) by ~ 10. However, even including these extra factors, the dual comb spectroscopy is hard pressed to compete in terms of sensitivity with a single swept cw laser spectrometer under identical conditions. The advantage of the dual comb system is the absolute frequency accuracy, high frequency resolution, and the broadband coverage that should allow for good discrimination against other absorbing species.

**IV.B Future Improvements**

We have shown that the stability of this system is sufficient for very long averaging times and that much of this averaging can be done in hardware to minimize the data burden. However, there are significant gains still to be made.

As discussed earlier increasing the dynamic range of the detector and digitizer could lead to an order of magnitude or more improvement in SNR. It is also very interesting to consider detecting the filtered signal in parallel, by use of a detector array, rather than sequentially. The system would be analogous to Fourier transform spectral interferometry [59, 60] with a similar increase in performance. In a dynamic-range limited system, the improvement is linear in the number of detector elements [16]. The rf heterodyne beat signals would have to be preserved and this might be possible through the use of a snapshot array. By setting $\Delta f_r$ to be small one could ensure that the highest rf beat frequency within the comb spectrum was below the readout speed so that no information was lost. Each snapshot element could then be processed in parallel to achieve significant improvement in averaging time, assuming the multiplicative noise could be sufficiently minimized to reach the shot noise limit. It is also interesting to note that the rf tones recorded on each element would still provide the frequency calibration of the measurement without a separate detector array calibration, a limit for Fourier transform spectral interferometry systems [59].

Probably the largest outstanding issue with future performance improvements remains with the comb sources themselves. Much of the recent development of frequency combs has been in support of frequency metrology and optical clocks [23, 24]. In the clock application, one typically only uses a few of the comb lines and requires high frequency accuracy over long times [14]. Dual-comb spectroscopy puts new and different requirements on the comb sources because it uses all of the comb teeth and requires high residual phase coherence between the two combs on short time scales (but generally lower absolute frequency accuracy). As a result, it is much more important to generate combs with strong, spectrally flat, phase coherent spectra than are used for clock applications. EOM based frequency combs [61, 62] offer a promising path to suppress phase noise but the flat spectra remains a challenge. Moreover, to realize the full potential of this approach, one would like these spectra to extend well beyond the gain region of available laser sources and into the mid-infrared, long wave-infrared and THz regions [1, 21]. There have certainly been advances in all these areas, but much more source development is needed to support dual comb spectroscopy, particularly in view of the usual tradeoffs faced between expanding the spectral coverage and retaining high levels of phase coherence and low relative intensity noise [63-65].

## V. Conclusion

Coherent dual comb spectroscopy can be viewed as massively parallel heterodyne laser spectroscopy, infrared time-domain spectroscopy, or a form of dispersive Fourier transform spectroscopy. We provide a general discussion of some of the important attributes of dual comb spectroscopy including frequency accuracy, resolution, normalization, and sensitivity. These considerations motivated our particular dual comb spectroscopy design, which is described in detail. We demonstrate that reasonably high SNRs can be achieved with a single point detector over a large number of resolved frequency elements through coherent averaging. Detector arrays or selected filtering bring the possibility of more rapid acquisition of high SNR spectra. We note many of the SNR and accuracy considerations here apply as well to dual comb LIDAR [66]. As a demonstration of the capabilities of dual comb spectroscopy, we measured the fully normalized, complex response of the ro-vibrational band of HCN for the first overtone of the C-H stretch. This measurement covers a 9 THz bandwidth with 41,000 resolution elements. The peak frequency domain uncertainty is 0.025 % in magnitude giving a peak fractional absorption sensitivity of 0.05% (converting to intensity) and 250 microradians in phase.

There are many other potential configurations of dual comb spectrometers including the incorporation of remote fiber delivery, long interrogation paths, multipass cells or cavities [8], detector arrays etc. It is also intriguing to consider combining this approach with multi-dimensional spectroscopy [67-69].

## Acknowledgements

We acknowledge useful conversations with Esther Baumann, Fabrizio Giorgetta, Todd Johnson, and Paul Williams and funding from the Department of Homeland Security and NIST.